# Research and Design of a Financial Intelligent Risk Control Platform Based on Big Data Analysis and Deep Machine Learning

Shuochen Bi[1,a*],Yufan Lian[2,b],Ziyue Wang[3,c]

{[a*]Email:bi.shu@northeastern.edu

[b]Email: yfanlian@126.com

[c]Email:Zw2013@nyu.edu}

[1]Independent Researcher D'Amore-McKim School of Business at Northeastern University Boston, MA02115,United States

[2]Independent Researcher D'Amore-McKim School of Business at Northeastern University Boston, MA02115,United States

[3]Independent Researcher，New York, NY 10012, United States

**Abstract:**In the financial field of the United States, the application of big data technology has become one of the important means for financial institutions to enhance competitiveness and reduce risks. The core objective of this article is to explore how to fully utilize big data technology to achieve complete integration of internal and external data of financial institutions, and create an efficient and reliable platform for big data collection, storage, and analysis. With the continuous expansion and innovation of financial business, traditional risk management models are no longer able to meet the increasingly complex market demands. This article adopts big data mining and real-time streaming data processing technology to monitor, analyze, and alert various business data. Through statistical analysis of historical data and precise mining of customer transaction behavior and relationships, potential risks can be more accurately identified and timely responses can be made. This article designs and implements a financial big data intelligent risk control platform. This platform not only achieves effective integration, storage, and analysis of internal and external data of financial institutions, but also intelligently displays customer characteristics and their related relationships, as well as intelligent supervision of various risk information.

## 1. Introduction

In the current US financial market, despite the rapid development of digital technology, the market still faces a series of challenges and difficulties. Against the backdrop of intensified inflationary pressures, the US financial market lacks sustained upward momentum and conditions. Compared with the rapid development of technology, traditional financial risk management models and warning methods have become increasingly outdated and insufficient, posing severe challenges and risks to financial institutions. Although sometimes the performance of financial markets may temporarily deviate from the macroeconomic situation, in the long run, the relationship between the market and economic development remains inseparable. Considering the current economic situation, the Federal Reserve is cautious about whether it has achieved victory in fighting inflation, and the possibility of raising interest rates has always existed. In many industries, except for some high-tech enterprises, the operating conditions of most enterprises are not as expected. The majority of recent stock index performance has been contributed by high-tech companies, and technology investment has also shown a downward trend. According to data from the National Venture Capital Association (NVCA), the investment amount and quantity of venture capital in the United States decreased significantly in the third quarter of this year. In this situation, the fundraising situation is also concerning, and it is expected that the annual fundraising amount will reach a new low in recent years. The US financial market faces many challenges, including economic cycles, inflationary pressures, and a decline in technology investment. Therefore, how to use emerging technologies to strengthen financial risk management and improve early warning capabilities has become one of the urgent problems to be solved in the current financial industry. In this context, this article will explore how to use advanced technological means to build a data-driven intelligent risk management platform based on the actual business needs of American commercial banks, in order to enhance the risk management capabilities and market competitiveness of financial institutions.

## 2.related research

### 2.1. Natural Language Processing (NLP) Big Data Technology

Natural Language Processing (NLP) is an important branch of artificial intelligence aimed at enabling computers to understand and process natural language like humans, achieving natural and smooth interaction between humans and machines.In his article 错误!未找到引用源。, R Mushtaq used a subfield of artificial intelligence (AI), natural language processing (NLP), to predict emotions. The research results indicate that the company's financial performance indicators help reduce negative emotions in the 10-ks text section. B Parker et al. explored the application of the state-of-the-art NLP model (BART) in their article and explored the use of data augmentation and various fine-tuning strategies to adjust it to optimal performance **错误:未找到引用源。**.Y Gao et al. proposed an adaptive sparse transformer Hawkes process (ASTHP) in the paper **错误:未找到引用源。**. By optimizing neural network parameters and using the relationships between events in the data to predict the type and occurrence time of the next event.

## 2.2. Big Data Technology and Machine Learning

Big data technology refers to the technology and methods for processing, storing, and analyzing large-scale, structurally complex, and rapidly growing datasets. Z Chen proposed a new model in the article for analyzing consumer data acquisition under privacy regulations **错误:未找到引用源。** and found that GDPR activates the data acquisition market by imposing consent requirements on data collection. Further research has been conducted on the optimization design of consumer data acquisition mechanisms, providing important policy insights for implementing social optimization. C Luo mentioned in his article **错误:未找到引用源。** that attribute reduction is an extremely important data preprocessing technique in today's data mining field, and proposed a parallel neighborhood entropy attribute reduction method based on neighborhood rough sets. This method utilizes the Apache Spark cluster computing model to parallelize algorithms in a distributed computing environment. Herlan studied the use of machine learning in his article to predict the reputation of customers during the renovation financing process **错误:未找到引用源。**. To evaluate alternative financing for energy conservation in Indonesia using machine learning and lifecycle cost analysis (LCCA). This model is built on the logistic regression model and artificial neural network model of machine learning. Developed and tested the model using Python algorithms, and demonstrated the efficiency of the proposed model. J Zhang studies the role of machine learning algorithms in the dynamic audit of Internet finance **错误:未找到引用源。**. Apriori algorithm, data mining, Bayesian network and other methods are proposed, and relevant experiments are carried out on the dynamic audit of Internet financial risks based on machine learning algorithm. The experimental results show that in Internet finance, private enterprises have the highest profit rate, up to 20%. But the higher the profit, the greater the financial risk.

## 2.3. Knowledge Graph Technology

Knowledge graph technology is a technique that utilizes methods such as graph theory and semantic networks to organize large amounts of data and information into semantically related knowledge structures, in order to achieve advanced information processing functions such as intelligent search, data mining, and semantic inference. Z Wang et al. mentioned in the article that Knowledge Graph (KG) is increasingly seen as an important resource in financial applications such as risk control, auditing, and anti fraud **错误:未找到引用源。**.

Multiple relation circles were introduced and a new embedding model was proposed, which takes into account the entity weights calculated by the PageRank algorithm to improve the TransE method. In the article 错误;未找到引用源。, Z Zhang considered bank data graphs and proposed a new HIDAM model for this purpose. Attempt to model commercial bank service scenarios by combining heterogeneous information networks with rich attributes on multiple types of nodes and links. To enhance the feature representation of MSE, we extract interaction information through meta paths and fully utilize path information. A Bouguerra used a knowledge-based perspective (KBV) in her article 错误;未找到引用源。 to investigate how and when companies in emerging economies can generate greater performance from absorptive capacity (AC).

## 3. Method

### 3.1. Functional module diagram of intelligent risk control platform

The big data intelligent risk control platform is a comprehensive application platform, and its various functional modules are closely connected, forming an organic whole. The data collection module is responsible for collecting a large amount of financial related data from various sources, including customer transaction information, market situation data, etc. These data are then transmitted to the data storage and management module, which is classified, stored, and labeled for subsequent retrieval and management. Supported by the data storage and management module, the intelligent mining and analysis module plays a crucial role by applying various algorithms to deeply mine and analyze data, constructing customer relationship models, risk models, etc. These models can not only be used for risk prediction and early warning, but also provide important data support for specific applications of the platform. Finally, the platform visualization and specific application docking module visualizes the analyzed data and intelligently pushes the results to relevant users, while outputting data reports for reference. As shown in Figure 1.

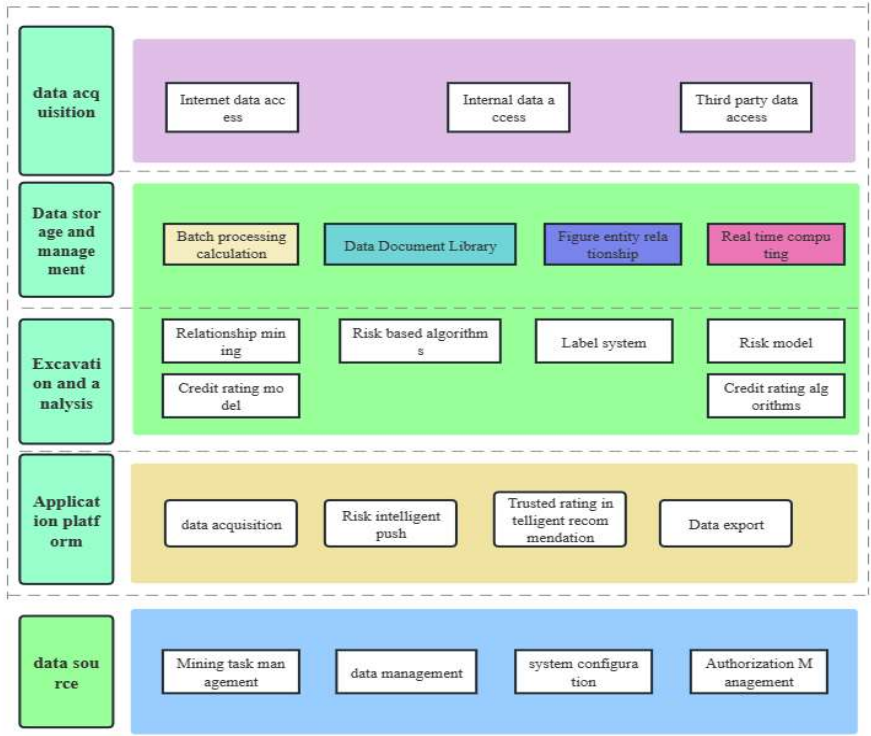

Figure 1    Functional module diagram of intelligent risk control platform

## 3.2.  Logic architecture diagram of financial big data risk control platform

The logical architecture of the financial big data intelligent risk control platform includes key modules such as data collection and storage, big data mining and analysis, and big data application layer. The data collection module is responsible for efficiently accessing various heterogeneous data and transmitting it to the data storage module for classification and management. The data storage module adopts multiple storage schemes to classify and level storage of different types of data, while supporting the import of various data sources, achieving unified management of massive data. The data processing module ensures data quality. In the big data mining and analysis module, the platform has built a stream processing parallel computing framework based on HDFS, effectively supporting sparse computing and iterative algorithms related to machine learning, providing strong technical support for risk assessment and prediction. The big data application layer enables historical data queries, batch processing applications, and marketing rules and models, providing risk management and business decision support for enterprises. The overall architecture connects the collection, storage, processing, analysis, and application of data, providing financial institutions with a complete set of intelligent risk management solutions. As shown in Figure 2.

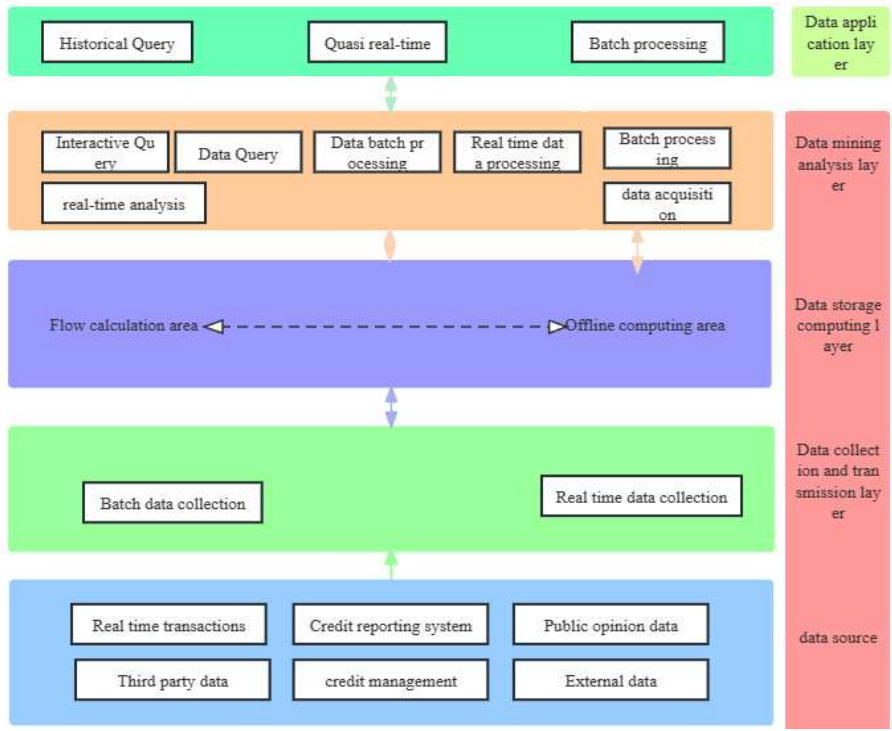

Figure 2    Logic architecture diagram of financial big data risk control platform

### 3.3.  Logic diagram of intelligent risk control model for financial big data

The logic diagram of the intelligent risk control model for financial big data includes key modules such as data warehouse construction, data analysis and mining, risk control warning and prediction, and visual display. Establish a financial data warehouse, unify the management of internal and external data, and provide services for subsequent analysis. Using data mining algorithms to analyze customer behavior characteristics, constructing a risk calculation library, and drawing customer behavior portraits. Using machine learning algorithms to construct risk control warning and prediction models, continuously improving their intelligence and accuracy. In the software design phase, implement the development of intelligent risk control platform functions and closely integrate the model with the software system. Realize automated early warning and visual display of financial risks, providing intelligent support for risk management. This logic diagram organically connects data establishment, analysis, prediction, and display, constructing a complete financial big data intelligent risk control model, providing strong risk management and decision support for financial institutions, as shown in Figure 3.

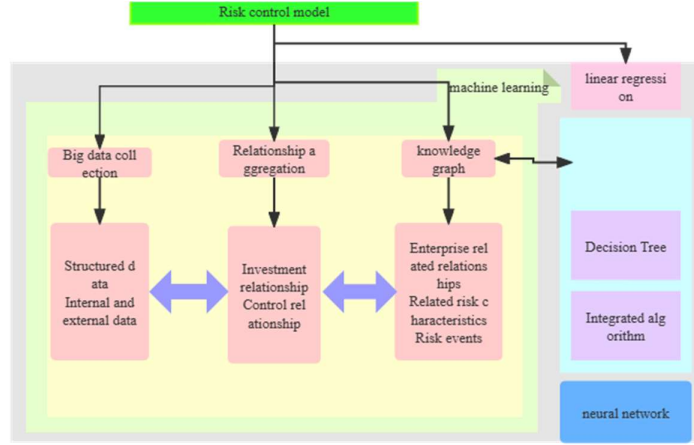

Figure 3 Logic diagram of intelligent risk control model for financial big data

The venture capital model plays a crucial role in risk assessment, investment decision-making, identifying investment opportunities, and optimizing investment portfolios by analyzing big data. It helps venture capital firms comprehensively evaluate the potential risks of investment projects and provides decision support to reduce investment risks. At the same time, venture capital models can also timely detect emerging trends and potential investment opportunities in the market, helping institutions grasp market changes.

## 4. Results and discussion

The platform testing environment configuration is shown in Figure 4.

| resource | application service | operating system | database | middleware | test tools |
|---|---|---|---|---|---|
| describe | 6CPU、32G | Suse11 | Oracle,hbase | Tomcat-8.5.9 | LoadRunner 11 |
| Remarks | virtual machine | | | | |

Figure 4 Test environment configuration

The main transaction load test is shown in Figure 5.

| Transaction Name | Enterprise related relationships | Customer credit rating | Multidimensional customer characteristics | Credit risk information |
|---|---|---|---|---|
| Duration | 10M | 10M | 10M | 10M |

| TPS value (pen/s) | 196.54 | 197.25 | 198.54 | 197.53 |
| response time | 1.274 | 1.465 | 1.421 | 1.396 |
| Transaction success rate | 100% | 100% | 100% | 100% |

Figure 5 Key Transaction Load Test Table

According to the above test data, all selected typical transactions performed well in terms of business processing time, all completed within 1500 milliseconds. This means that the system response time remains within 3 seconds, meeting our strict requirements for system performance. The utilization rate of various system resources is within the expected range, and memory applications and releases are proceeding normally. The overall remaining memory remains stable and there are no performance issues such as memory leaks or abnormal crashes.

After conducting load verification tests on the key functional points of the platform, we found that the maximum capacity TPS of each transaction interface is about 200, which is in line with our performance testing goals. When conducting high concurrency verification tests, we noticed that the overall running status of the platform was good, and there were no obvious performance issues or system crashes. In addition, we have also observed that the physical resource consumption of the platform is within an acceptable range, which means that the platform can still maintain stability in handling high concurrency situations, which is crucial for ensuring the reliability and availability of the platform.

Compared to the capital market, the performance of the US money market is stable. In May 2023, the size of the US commercial paper market was $1.18 trillion, which is almost the same as the $1.18 trillion in May 2021 and $1.17 trillion in May 2022. Overall, the direct financing function of the US capital market has not been fully utilized, which means that the future financing demand of enterprises is relatively weak. Compared to normal years, stock market financing is still in a state of suppression, but the recovery of bond market financing is slightly faster. In the large economic environment both domestically and internationally, enterprises tend to adopt a wait-and-see attitude and will not rashly add investment. Although financing activities in the capital market have not fully resumed, this environment also provides valuable opportunities for risk control platforms. We also need to strengthen intelligent applications while ensuring system security, which can better meet the personalized needs of customers and improve service levels. In addition, cross platform adaptation is also an important direction for future development to ensure seamless connection of financial services on different terminal devices. In summary, future prospects include further optimizing performance, expanding functionality, strengthening intelligent applications and cross platform adaptation, in order to address the challenges of the financial market and provide customers with more comprehensive and efficient financial services.

## 5、Conclusion

In the first two years after the outbreak of the COVID-19 epidemic (2020-2021), as residents and enterprises sharply cut spending, increased savings and received a lot of policy support, the balance of deposits of U.S. banking institutions hit new highs. In this environment, many local banks need to find ways to increase interest income. In the context of the current US financial market, with the vigorous development of Internet finance, cloud computing, big data, artificial intelligence and other cutting-edge technologies, the financial industry is undergoing unprecedented changes. This study closely focuses on this trend and delves into the key issues of intelligent risk control in banks under the background of financial big data. By constructing a financial big data intelligent risk control platform and combining it with the actual situation of American commercial banks, this study aims to provide important theoretical and practical guidance for the banking industry to respond to challenges, improve risk management level, and achieve intelligent development in the digital era. This platform can not only respond to the dynamic changes in the financial market, but also better meet the financing needs of American enterprises and customers, promoting continuous transformation and innovation of the financial system. Therefore, this study provides useful references and insights for the US banking industry in adapting to new technological trends, improving competitiveness, and achieving sustainable development..

## References:

[1]Mushtaq R , Gull A A , Shahab Y ,et al.Do financial performance indicators predict 10-K t ext sentiments? An application of artificial intelligence[J].Research in International Business and Finance, 2022, 61.DOI:10.1016/j.ribaf.2022.101679.

[2]Parker B , Sokolov A , Ahmed M ,et al.Domain Specific Fine-tuning of Denoising Sequenc e-to-Sequence Models for Natural Language Summarization[J]. 2022.DOI:10.48550/arXiv.2204.0 9716.

[3]Gao Y , Liu J W .Adaptively Sparse Transformers Hawkes Process[J].International journal of uncertainty, fuzziness and knowledge-based systems: IJUFKS, 2023(4):31.DOI:10.1142/S0218488 523500319.

[4]Chen Z .Privacy Costs and Consumer Data Acquisition: An Economic Analysis of Data Priv acy Regulation[J].Monash Economics Working Papers, 2022.

[5]Luo C , Cao Q , Li T ,et al.MapReduce accelerated attribute reduction based on neighborho od entropy with Apache Spark[J].Expert Systems with Application, 2023.

[6]Herlan, Sudarmaji E , Yatim M R .Predictive Creditworthiness Modeling in Energy-Saving F inance: Machine Learning Logit and Neural Network[J].Financial Risk and Management Review s, 2022, 8.

[7]Zhang J .Dynamic Audit of Internet Finance Based on Machine Learning Algorithm[J].Mobil e information systems, 2022(Pt.24):2022.

[8]Wang Z , Yang L , Lei Z ,et al.An entity-weights-based convolutional neural network for large-sale complex knowledge embedding[J].Pattern Recognition: The Journal of the Pattern Recognition Society, 2022.

[9]Zhang Z , Ji Y , Shen J ,et al.Heterogeneous Information Network based Default Analysis on Banking Micro and Small Enterprise Users[J].Papers, 2022.DOI:10.48550/arXiv.2204.11849.

[10]Bouguerra A , Mellahi K , Glaister K ,et al.Absorptive capacity and organizational performance in an emerging market context: Evidence from the banking industry in Turkey[J].Journal of Business Research, 2022, 139.